\documentstyle[12pt,aasms4]{article}

\def\bull{\vrule height .9ex width .8ex depth -.1ex } % square bullet

\lefthead{Beaulieu et al.}
\righthead{Dynamics of the Galactic Bulge}

\begin{document}

\title{Dynamics of the Galactic Bulge using Planetary Nebul\ae}

\author{Sylvie F. Beaulieu\altaffilmark{1}, Kenneth C. Freeman, Agris J. Kalnajs,
Prasenjit Saha\altaffilmark{2},}
\affil{Mount Stromlo and Siding Spring Observatories, The Australian National
University, Canberra ACT 2611, Australia; beaulieu@ast.cam.ac.uk; 
kcf@mso.anu.edu.au; agris@mso.anu.edu.au; saha@maths.qmw.ac.uk}

\altaffiltext{1}{present address: Institute of Astronomy, University of 
Cambridge, Madingley Road, Cambridge CB3 0HA, United Kingdom}
\altaffiltext{2}{present address: Astronomy Unit, School of Mathematical Sciences,
Queen Mary and Westfield College, Mile End Road, London E1 4NS, United Kingdom}

\author{and HongSheng Zhao}
\affil{Sterrewacht Leiden, Niels Bohrweg 2, 2333 CA, Leiden, Netherlands;
hsz@strw.LeidenUniv.nl}

\begin{abstract}

Evidence for a bar at the center of the Milky Way triggered a renewed enthusiasm 
for dynamical modelling of the Galactic bar-bulge. Our goal is to compare
the kinematics of a sample of tracers, planetary nebul\ae, widely distributed
over the bulge with the corresponding kinematics for a range of models
of the inner Galaxy. Three of these models are N-body barred systems arising from
the instabilities of a stellar disk (Sellwood, Fux and Kalnajs), and one is a
Schwarzschild system constructed to represent the 3D distribution of the COBE/DIRBE
near-IR light and then evolved as an N-body system for a few dynamical times
(Zhao). 
For the comparison of our data with the models, we use a new technique developed 
by Saha (1998). The procedure finds the parameters of each model, i.e. the
solar galactocentric distance $R_\circ$ in model units, the orientation angle 
$\phi$, the velocity scale (in km s$^{-1}$ per model unit), and the solar 
tangential velocity which best fit the data.

\end{abstract}


\section{Introduction}

A spiral galaxy consists of a relatively flattened stellar disk in nearly 
circular rotation and, in most systems, a central bulge. It is estimated that
about 30 \% of these galaxies also show a central bar in the visible; 
however the real fraction
of barred galaxies is probably significantly higher because some apparently normal
spirals show a bar feature in the near-IR that was not visible in their optical
images (e.g. Sellwood \& Wilkinson, 1993). In addition, barred galaxies often show
a lens and/or ring around the bar. The flattened disks contain objects of all ages,
from the interstellar gas and very young stars to the old disk stars which in our 
Galaxy are almost as old as the globular clusters. The bulges appear to be made up
mainly of old stars.

The disks of most disk galaxies are relatively thin, with the ratio of their
radial to vertical scale heights mostly in the range 5 to 15. In the later-type 
barred galaxies, the central bar may be no thicker than the host disk. Kormendy 
(1993) has argued that many of the features identified as bulges from the surface
photometry of more face-on galaxies may also be as thin as the disks. However,
many edge-on galaxies show bulges which clearly do extend beyond the disk.

The bulges of spiral galaxies show a wide range of shapes, from spheroidal
through boxy or peanut shaped bulges. The boxy versus spheroidal 
structure of bulges is roughly understood in terms of their orbital
properties but not in terms of origin. Many possibilities have been
suggested for the origin of boxiness in bulges, including the formation and
dissolution of bars, dissipative processes during the collapse of a rapidly
rotating inner region, or later accretion events (see Sellwood 1993;
Rowley 1986; Whitmore \& Bell 1988; Combes et al. 1990; Pfenniger et al. 1991).

Kormendy and Illingworth (1982) pointed out that the boxy bulges are frequently
cylindrical rotators, unlike the more spheroidal bulges. This led to a burst of
observational and theoretical studies of these systems (e.g. Binney \& Petrou 1985; 
Rowley 1986; Shaw 1993), with the growing indication that these boxy or
peanut-shaped edge-on systems may be associated with bars (Combes et al. 
1990; Sellwood \& Wilkinson 1993).

The Milky Way has an excellent example of a box-shaped bulge. This feature 
was seen in the early $2.4\mu$m balloon scans (Matsumoto et al. 1982),
and spectacularly confirmed by the $2.2\mu$m image of the Galaxy from 
COBE\footnote{Cosmic Background Explorer}/DIRBE\footnote{Diffuse Infrared 
Background Experiment}$^{,}$\footnote{The COBE datasets were developed by 
NASA Goddard Space Flight Center under the guidance of the COBE Science 
Working Group and were provided by the NSSDC.} (Weiland et al. 1994; 
Arendt et al. 1994) as seen from the contours plot of the COBE/DIRBE 2.2$\mu$m 
image (Figure 1).
See Binney et al. (1997) for a dust-corrected non-parametric recovery of
the light distribution in the inner few kpc of the Milky Way from the
COBE/DIRBE surface brightness map.

The bulge of the Milky Way provides a unique opportunity
to investigate the detailed pattern of rotation and velocity dispersion in a
boxy Galactic bulge. We can study the structure of the bulge to see if
this boxy bulge is really a stellar bar, and we can also see how the bulge
and disk are related dynamically.

This paper is outlined as follows. We start with an overview of recent studies 
of the bar/bulge problem through axisymmetric and N-body models (Section 2).
In Section 3, we discuss the wide range of tracers available to study the
kinematics of the Galactic bulge, and the data obtained for this study.
A preliminary visual assessment of the data is presented in Section 4 
using the mean velocity and velocity dispersion versus the Galactic longitude
and latitude. Section 5 compares our planetary nebul\ae~ (PNe) distribution with 
the distribution of light in the COBE/DIRBE images in the 1.25, 2.2 and $3.5\mu$m
wavelength regions. In Section 6, we compare our PNe data with four Galactic
bar-bulge models: three are N-body models and one is a relaxed Schwarzschild 
realization
of the COBE light distribution. A summary and conclusions are given in Section 7. 

\section{Dynamics of the Bulge}

\subsection{The Bar/Bulge of the Milky Way}

Evidence is accumulating that the boxy peanut-shaped bulges seen in edge-on disk 
galaxies are associated with bar structures (Combes et al. 1990; Jenkins \& 
Binney 1994; Blitz \& Spergel 1991; Kuijken \& Merrifield 1995; Bureau \&
Freeman 1999). For the bulge of our Galaxy, the $2.4\mu$m balloon 
scans and the near-IR COBE/DIRBE images show
such a boxy peanut shape. An unambiguous direct identification of a bar at the 
center of the Galaxy is difficult because the Sun is located in the plane of
the Galaxy and our view of the Galactic center is obscured by the dust.
The patchy extinction in the plane of the Galaxy is
obvious from the optical image of the Galactic bulge taken
at ESO (Madsen \& Laustsen, 1986). It is clear that the southern part of the
bulge is much less affected by extinction than the northern part. The
southern part includes two famous regions of relatively low extinction,
Sgr I ($l= 1.4^\circ$, $b = -2.6^\circ$) and Baade's Window (BW)
($l = 1.0^\circ$, $b = -3.9^\circ$)
which are widely used for studies of the stellar population and dynamics of the
inner bulge. The distribution of extinction over the bulge is also nicely shown
from the work of the COBE/DIRBE group (Arendt et al. 1994, figure 3b,
plate L7).

Nevertheless, much observational evidence is now pointing to the existence of
such a bar. Here, we list only a few: see Gerhard (1999) for a more detailed 
review.

\begin{itemize}
\vspace{-2mm}
\item de Vaucouleurs (1964) was the first to point out that a central bar is 
probably responsible for the non-circular motions of the HI in the inner part 
of the Milky Way. He had already noted that similar non-circular motions were 
present in the inner parts of barred spiral galaxies. 
\vspace{-2mm}
\item The asymmetry in longitude of the distribution of the $2.4\mu$m emission 
derived from the balloon scans indicated that the stars in the 
central kpc lie in a bar with its near side at positive Galactic longitude 
and suggested that the bar is tilted relative to the Galactic plane
(Blitz \& Spergel 1991).
\vspace{-2mm}
\item The COBE/DIRBE images (Weiland et al. 1994) confirmed the asymmetry
in the surface brightness distribution of the bulge in the near-IR, but show
no evidence for an out-of-plane tilt of the bar.
\vspace{-2mm}
\item Nikolaev \& Weinberg (1997) reported that the distribution of variables in the 
IRAS Point Source Catalogue (PSC)
is consistent with a bar with semi-major axis of 3.3 kpc and position angle of 
24$^\circ~ \pm 2^\circ$ (where position angle is the angle between the major
axis of the bar and the Sun-center line and is taken as positive for a bar 
pointing into the positive Galactic longitude quadrant).
\vspace{-2mm}
\item Rohlfs and Kampmann (1993) showed that the HI terminal velocities indicate 
the presence of a bar with a semi-major axis of $2-3$ kpc and a position angle of 
about 45$^\circ$.
\vspace{-2mm}
\item Binney et al. (1991) used CO kinematics in the inner parts of the
Galaxy to show the presence of a bar with a pattern speed of 63 
$\rm km s^{-1} kpc^{-1}$, a corotation radius of 2.4 kpc and a position angle of 
$16^\circ \pm 2^\circ$. More recent gas dynamical studies (Englmaier \& Gerhard 
1999; Weiner \& Sellwood 1999; Fux 1999) all support for a substantially larger 
corotation radius.
\end{itemize}
 
From some of these studies, and others on the brightnesses of tracer objects like
Mira variables (e.g. Whitelock 1993) and clump giants in the bulge 
(e.g. Stanek et al. 1994), it seems 
fairly clear that the bulge objects at positive Galactic longitude are brighter
than those at negative longitude. This is generally interpreted as evidence that 
we are viewing the bar/bulge at an angle from its major axis and that the closer
end of the bar is at positive longitude. There is still disagreement on the
parameters of the bar, i.e. its length, strength, pattern speed and position angle.
But, if we were to view the Galaxy edge-on from outside, it would probably
look much like NGC 891, with probably more bulge than NGC 891 but less than
NGC 4565 (see the Hubble Atlas).

\subsection{Axisymmetric models}

Kent (1992) used infrared ($2.4\mu$m) surface photometry from
the Spacelab infrared telescope to make an axisymmetric model for the
luminosity density distribution in the inner galaxy. For the disk, he
modelled the luminosity density $L$ as a double exponential in $R$ and $z$,
and for the bulge he adopted
$$L(R,z) = 3.53 K_\circ(s/667)~~~L_\odot~\rm pc^{-3}~~~{\rm for}~s>938$$ 
and $$L(R,z) = 1.04 \times 10^6 (s/0.482)^{-1.85}~~~L_\odot~\rm pc^{-3}~~~{\rm for}~s<938.$$

\noindent
where $K_\circ$ is a modified Bessel function.
Here $s^4 = R^4 + (z/0.61)^4$ and the units of s in the equation above
are parsecs. This form of the $L(R,z)$ distribution for the bulge leads
to box-shaped isophotes.

Kuijken (1995) used a quadratic programming technique on a
bilinear tessellation in the energy, angular momentum $(E,L)$ plane to construct
a two-integral distribution function $f(E,L)$ for a slightly modified version
of Kent's axisymmetric model for the inner Galaxy. The distribution function
is forced to give an isotropic velocity dispersion. With Kent's values for the
mass to light ratios for the disk and bulge, the predicted line-of-sight velocity
distribution in Baade's window is in excellent agreement with the distribution
observed for the M giants by Sharples et al. (1990). However, the agreement is
not so good for the velocity distribution of the K giants in Minniti's (1992)
field at $l = 8^\circ,~b = 7^\circ$:
the discrepancy between the data and the prediction from the
distribution function $f(E,L)$ is seen in the mean velocity and in the
shape of the velocity distribution in this region. Kuijken suggests
that the discrepancy might be associated with the triaxiality of the
bulge, and points out how remarkable it is that his oblate, isotropic
and axisymmetric model gives such a good fit to the velocity distribution
in Baade's Window.

Durand et al. (1996) used a two-integral axisymmetric model with a
Kuzmin-Kutuzov St\"{a}ckel potential (with a halo-disk structure) to 
study the dynamics of a sample of 673 PNe taken 
from the Acker et al. (1992) catalogue. The method fits the kinematics
to the projected moments of a distribution function by means of Quadratic 
Programming. They conclude that their two-integral model does not adequately 
characterize the dynamical state of their sample of PNe.

Our particular interest here is in investigating the triaxial structure of the
bar-bulge further, so we will not pursue the axisymmetric models in this paper. 
The question is about the origin of central bar-bulges: do they arise from 
instabilities of the disk of galaxies or from other processes like the accretion 
of satellites or as part of the dissipative collapse of the galaxies ? The
quantitative study of the formation of bars through disk instabilities is now
well advanced through N-body models, which we now discuss briefly. In \S 6.3,
we will compare the kinematical properties of our PNe and the models. For the
N-body models, this comparison will show whether the instability picture gives
a plausible description of the observed bulge kinematics.

\subsection{N-body models}

In the last few years, the growing evidence for a bar at the center of our Galaxy 
initiated much interest in developing detailed dynamical models of the
Galactic bar-bulge. Different kinds of models are now available, but the
observational constraints on their stellar dynamics are not yet well advanced.

N-body models of the bar-forming instabilities of disks provide theoretical
predictions of the dynamics of the resulting bar-bulges which can be tested
against dynamical data from the Galactic bulge and other bulges. For example,
Fux (1996), Sellwood (1993) and Kalnajs (1996) have all modelled the central
bar-bulge through the instabilities of self-gravitating stellar disks. 
As tests of the relevance of these models to the dynamics of the Galactic bulge,
the detailed kinematics of their models can be compared with the observed
kinematics of tracer bulge objects like the PNe which are the
subject of this paper.

Another kind of numerical model for the Galactic bulge comes from the work of
Zhao (1996) who constructed a rotating Schwarzschild model for the COBE light
distribution. Although this model does not provide direct insight into the
formation of the bulge, in the way that the studies of disk instabilities can
do, the Schwarzschild model is of much interest for evaluating the present
dynamical state of the bulge. For this purpose, we can compare the kinematics 
of N-body realizations of this model with observational data, as above.
 
It would be most desirable if we could obtain an unbiased spatial distribution
and the radial velocities of a subset of bulge objects. Such a database
would allow us to distinguish between the various proposed models, and no doubt 
suggest others. Unfortunately most of the stellar objects have to contend with 
the high and patchy absorption near the Galactic plane [OH-IR stars are a clear
exception].

\section{Planetary Nebul\ae \, as tracers}

To study the kinematics of the Galactic bulge, we have access to a wide range
of tracers: OH/IR stars (Habing 1993; Sevenster et al. 1997a,b), Miras 
(Whitelock 1993), M giant stars, K giant stars (both individually and through
the integrated bulge light) (Walker et al. 1990; Minniti et al. 1992; Minniti 
1996a,b; Terndrup 1993; Ibata \& Gilmore 1995a,b), carbon stars (Whitelock 1993), 
SiO Maser sources (Deguchi 1997), RR Lyr\ae~ stars
(Walker \& Terndrup 1991) and PNe (Kinman et al. 1988; Durand et al. 1998).
The highly evolved OH/IR stars, Miras and M giants stars are probably biased
towards the metal-rich population: the radial distribution of these objects is
significantly steeper than the distribution of integrated light in the bulge
(e.g. de Zeeuw 1993), and the kinematics of these objects reflects the kinematics of
the metal-rich component of the bulge (Sevenster 1997). The carbon stars  
are rare and are also an indication of an intermediate age metal-rich population. 
The K giant stars are found at all metallicities and would be the ideal tracers to 
use since all bulge stars are likely to go through a K giant phase, but they are 
relatively faint. The K giants have already provided important dynamical 
information (e.g. Terndrup et al. 1995; Ibata \& Gilmore 1995a,b), and much
more will appear in the future from the large fiber surveys in progress (e.g. Harding
\& Morrison 1993). The RR Lyr\ae~ stars are also useful bulge tracers but they 
are biased toward the metal-poor population and are fainter than the K giants. 
The PNe are not biased towards the metal-rich population (e.g. Hui et al. 1993): 
recall the presence of PNe in the very metal-poor globular cluster M15 (Pease 1928). 
Their spatial distribution and their high velocity dispersion indicate that most 
of the bulge PNe are old objects. Their strong H$\alpha$ and [OIII] emission
lines make their velocities easy to measure. We have thus decided to use the
PNe as probes to study the kinematics and dynamics of the Galactic bulge. 

The distances of PNe are still poorly known. Using the optical diameter as a 
distance criterion is not adequate because PNe have a wide range of absolute 
diameters. Nevertheless, using the angular diameters, spatial distribution
and radial velocities of a sample of PNe, Gathier et al. (1983) estimated 
that probably 80\% of the small (diameter $< 20''$) PNe within $10^\circ$ of the
Galactic center belong to the bulge. While it is clear that most of the PNe towards 
the bulge are associated with the bulge, it is also evident that their apparent 
spatial distribution at low Galactic latitudes is affected by the interstellar
absorption.

\subsection{The Data}

In 1994 and 1995, we conducted an H$\alpha$ imaging survey 
of the Galactic bulge in order to detect new PNe (Beaulieu et al. 1999). The  
survey yield 56 new and 45 already catalogued PNe. We obtained radial velocities
for each new PNe plus a sample of 317 catalogued PNe (i.e. 272 catalogued and the
45 rediscovered PNe) taken from the Strasbourg-ESO Catalogue of Galactic Planetary 
Nebul\ae~ (Acker et al. 1992). Although we intended to observe only the southern 
part of the bulge (less affected by extinction), we have obtained a few fields
in the northern part as well.
Our data have already been used in a study of Galactic kinematics by Durand
et al. (1998).

Our database of PNe contains two samples. The first sample comprises the 97 PNe
(new and rediscovered) found in the southern bulge from our uniform survey
with the 1.0m telescope. The region covered by this survey is 
$-20^\circ < l < 20^\circ$ and $-5^\circ > b > -10^\circ$. We will 
refer to this uniformly selected sample as the {\it Survey fields only} sample.
A note is needed here: this sample, in fact, contains 98 PNe but we are using
97 PNe for the analysis. The reason for this is that we accepted one PN as
``probable" after we have completed the {\it Survey fields only} sample
analysis. This PN is SB15 : $\rm PN G009.3-06.5$.
 
The second sample is less homogeneous, with the 98 PNe {\it Survey fields only} 
sample (including, this time, {\it SB15}), the 3 PNe which we discovered in the
northern bulge, and the 272 PNe from the Acker et al. (1992) catalogue for 
which we have measured new radial velocities. This larger sample contains
373 PNe and covers the more extended region $-30^\circ < l < 30^\circ$ and
$3.3^\circ < |b| < 15^\circ$. We will refer to this sample as the
{\it Survey fields} + {\it Catalogue} sample. Figure 2 shows the ($l,b$)
distributions for the two samples.
 
In the absence of information on distances for our PNe, we made no
attempt to separate disk and bulge PNe in our two samples. Therefore,
disk contamination is likely. We note, however, that some of the
dynamical models used in this study (see \S 6) include a disk. 

\section {Analysis}

In the first part of this section, we present several plots showing the
kinematics of these two samples for preliminary visual assessment. 
We then go on to compare the properties of the PNe 
samples with the properties of several recent dynamical models. This
comparison will be first presented visually in the form of plots of
individual velocities, mean velocities and velocity dispersions against
$l$ and $b$. Then we will use a statistical technique by Saha (1998) to
make a more quantitative comparison of the data with the models, and
to estimate the Galactic scaling parameters and orientations which
best match the models to our data.

The typical radial velocity error for our PNe is 11 km s$^{-1}$ 
(Beaulieu et al. 1999). For the Galactic bulge, the velocity dispersion ranges
from about 60 km s$^{-1}$ to 125 km s$^{-1}$ (Fig. 13), so this radial
velocity error is negligible.

In the presentation of the kinematics of our samples, in order to illustrate
the systemic rotational properties of the bulge PNe more clearly,
we will show the velocities of the PNe corrected for the solar
reflex motion. We adopted the circular velocity of the Local Standard
of Rest (LSR) at the Sun as 220 km s$^{-1}$ (Kerr et al.
1986). For the Sun's peculiar velocity relative to the LSR we use
16.5 km s$^{-1}$ towards $l = 53^\circ$, $b = 25^\circ$ (e.g. Mihalas
\& Binney 1981). The corrected line-of-sight ($V_{los,GC}$) velocity 
(i.e. the line-of-sight velocity in km s$^{-1}$ that would be observed
by a stationary observer at the location of the Sun) is then given by

\[ {V_{los,GC} = V_{obs}} + 220 \sin l \cos b  + 16.5\,[\sin b \sin 25 + \cos b \cos 25 \cos(l - 53)] \]
 
\noindent
where $V_{obs}$ is the heliocentric observed line-of-sight velocity
in km s$^{-1}$.

Figure 3 shows the longitude versus velocity diagram for the 
{\it Survey fields only} (top panel) and {\it Survey fields} + {Catalogue}
(lower panel) (corrected for the solar reflex motion).

In the figures that follow, we note that there must be some level of distance 
bias in our PNe samples. The longitude distribution of the PNe shows some evidence
for depletion at $l<0$ (the more distant side of the bar) relative to $l>0$
(Fig. 14), although this depletion is only marginally significant (Fig. 16).
In the comparisons of the PNe distribution and kinematics with the various models
(\S 6), we will ignore this distance bias.  

\subsection{Survey fields only}

Figure 4 shows the longitude versus mean velocity (top panel) and the longitude
versus velocity dispersion (lower panel) 
using 8 bins in longitude, with approximately equal numbers of PNe in each bin 
(12 to 13 PNe). The rotation of the bulge is clearly seen, with an amplitude
of about $\pm 100$ km s$^{-1}$. The velocity dispersion of the bulge is
approximately constant with longitude, except for the apparent drop in
$\sigma$ for $l>+12^\circ$. This drop is
seen again in the larger sample described in \S 4.2 but on both sides of
the Galactic center: see Figure 7. It is probably due to the contribution 
of the inner disk at these longitudes (see Lewis \& Freeman 1989).

Figure 5 shows the latitude versus mean velocity (top panel) and the latitude
versus velocity dispersion (lower panel) for 2 bins with 
equal number of PNe in latitude. Each bin in latitude contains 48 to 49 PNe.
We see that the {\it total} velocity dispersion about the mean velocity 
does not appear to change significantly with latitude. [Note that this total
velocity dispersion in the plots against latitude includes the systemic
rotation and random velocities of the stars.]

Tables 1 and 2 summarize the binned data shown in Figure 4 and 5 
respectively. Column 1: the mean latitude and longitude, Column 2:
the mean velocity (km s$^{-1}$), Column 3: the velocity dispersion, Column 4:
the error (standard deviation) in the mean velocity, and Column 5: the error
(standard deviation) in the velocity dispersion. 

We have also divided the $l-V$ diagram of Figure 3 (top panel) into two bins in 
Galactic latitude (with 48 to 49 PNe in each bin) (Figure 6) in order to see if
contamination from disk PNe is affecting our data. Disk contamination is
potentially more serious at higher latitudes because of the steeper density
gradient of the bulge. Therefore, if contamination were present, we would expect 
the lower latitude bin ($b = -04.9^\circ$ to $-06.5^\circ$) to be 
significantly hotter (i.e. have higher velocity dispersion) than the higher 
latitude bin ($b = -06.6^\circ$ to $-10.2^\circ$). We see no evidence in
Figure 6 for serious disk contamination in our sample, except possibly for
$l>+12^\circ$.

\subsection{Survey fields and Catalogue objects}

Now we present the data for the larger and more extended but less 
homogeneous {\it Survey fields} + {\it Catalogue} sample of 373 PNe. 
(see Figure 3 (lower panel)). Figure 7 shows the longitude versus mean velocity (top
panel) and the longitude versus velocity dispersion (lower panel) using 12 bins in
longitude with approximately equal numbers (31 to 32) of PNe in each bin.
Again, the rotation of the bulge is clearly seen. For $|l| > 12^\circ$, 
the mean rotational velocity  continues to rise as the data become dominated 
by PNe of the inner disk. In this larger sample, beyond $|l| > 12^\circ$,
we see again an apparent drop in the velocity dispersion, due presumably
to the contribution of the inner disk PNe at these longitudes. 

Figure 8 shows the latitude versus mean velocity (top panel) and the latitude
versus {\it total} velocity dispersion (lower panel) using 6 bins in
latitude with approximately equal numbers (62 to 63) of PNe in each bin; 
2 bins are in the northern bulge and 4 bins in the southern bulge. [Note 
again that the total velocity dispersion in the latitude plots includes
the systemic rotation and random velocities of the stars.] 

Tables 3 and 4 summarize the binned data shown in Figure 7 and 8 
respectively. Column 1: the mean latitude and longitude, Column 2:
the mean velocity (km s$^{-1}$), Column 3: the velocity dispersion, Column 4: 
the error (standard deviation) in the mean velocity and Column 5: the error 
(standard deviation) in the velocity dispersion. 

In Figure 9, we are looking again at the disk contamination using the
same latitude bins as for the {\it Survey fields only} sample. In this
larger and more extended sample, contamination from the disk PNe becomes 
evident outside the 
longitude region $|l| = 12^\circ$ where the PNe velocity distribution
becomes significantly colder.

We also present a series of longitude-velocity diagrams for 6 bins in
latitude. Figure 10: $b = +03.3^\circ$ to $+05.2^\circ$ (top panel)
and $b = +05.2^\circ$ to $+15.1^\circ$ (lower panel). Figure 11: 
$b = -03.3^\circ$ to $-04.4^\circ$ (top panel) and 
$b = -04.5^\circ$ to $-05.8^\circ$ (lower panel). Figure 12:
$b = -05.8^\circ$ to $-07.4^\circ$ (top panel) and
$b = -07.4^\circ$ to $-14.9^\circ$ (lower panel).
For this less homogeneous (and generally brighter) sample of PNe,
the disk contamination really starts to show in the two high latitude bins: 
the PNe velocity dispersion becomes much colder at all longitudes, as we would 
expect to see if the disk contamination is significant at higher latitudes. 

\subsection{Comparison with other studies}

In recent years, there have been some important studies of the kinematics of
K and M giants in the Galactic bulge. Although the regions observed are mostly 
not as extended as our survey, we should now compare the kinematics
derived from these studies with the results from the PNe. 

In Figure 13, we show again the mean velocity and velocity dispersion 
against longitude for our extended sample, and have overplotted data from 
kinematic studies of giants, which fall in our {\it Survey fields} + 
{\it Catalogue} sample region. Minniti (1996a) presented data for
three bulge fields. He gives kinematical data for the more metal-rich 
([Fe/H] $> -1$: filled symbols) and metal-poor stars ([Fe/H] $< -1$: open symbols) 
separately. Data for one field come from Harding and Morrison (1993), and 
again we show the data points for the more metal-rich and more metal-poor 
stars separately. For Baade's Window (K giants: Terndrup et al. 1995), we show
the only available data point, the velocity dispersion value, for his stars with 
V $\geq$ 16.0: these fainter stars are likely to be a relatively uncontaminated
sample of bulge stars. Sharples et al. (1990) find an almost identical dispersion 
for their M giants in Baade's Window. Finally, we present three data points 
(higher latitude [$b=-12^\circ$]) 
from Ibata \& Gilmore (1995a,b). We derived equivalent $<V_{los,GC}>$
values for their three negative longitude fields from the gradients $\Omega_G$
that they estimated assuming an isotropic velocity dispersion. We used the
formalism of Morrison et al. (1990), assuming that the stars in each field
lie where the line-of-sight passes closest to the center of the bulge.
Ibata and Gilmore give kinematical solutions for several assumptions about
the shape of the bulge velocity ellipsoid $\sigma$. The derived $<V_{los,GC}>$
values depend very weakly on the assumptions about the shape of $\sigma$, so we
have only plotted the isotropic solution (asterisks) in Figure 13 (upper panel).
Their velocity dispersions are more sensitive to the shape of $\sigma$. We show
their velocity dispersions for an isotropic bulge (asterisks) in Figure 13 (lower
panel). The isotropic solution for $\sigma$ appears to give better agreement 
of the Ibata and Gilmore data with the other bulge samples.  
For comparison we also show (line) the slope of the linear rotation
curve found for 279 bulge PNe by Durand et al. (1998). (Part of our data is
included in their analysis.) The slope of this line is 
9.9 $\rm km~s^{-1}~degree^{-1}$. Table 5 summarizes 
the symbols associated with each study. Column 1: the study, Column 2: the 
field's ($l,b$) coordinates, and Column 3: the symbol used on the plot.

For Minniti's three fields, the data for the metal-rich giants clearly matches
our PNe data better than do the metal-poor giants. For the Harding-Morrison 
field, although we see the same match of the metal-rich giants with our PNe in the
velocity dispersion, it is in fact the opposite that is seen for the mean
velocity. This disagreement was also observed by de Zeeuw (1993) when he
compared the Minniti and Harding-Morrison samples with Kent's model (Kent 1992). 
Nevertheless, the otherwise good agreement seen so far identifies the bulge 
PNe with 
the more metal-rich giant ([Fe/H] $> -1$) of the bulge, as we would expect. For
Baade's Window, it is interesting to see that the velocity dispersion is perhaps
somewhat higher than the mean of the velocity dispersion values for
our PNe at lower $|l|$, but we note that our PNe are mostly more distant
from the Galactic plane than Baade's Window (cf Figure 8). (The velocity
dispersion along the minor axis of the Galactic bulge is known to decrease 
with increasing $|b| \gtrsim 2^{\circ}$: e.g. Rich 1996.) 

We note that the $<V_{los,GC}>$ values shown in Figure 13 for the Ibata and
Gilmore sample pertain to their more metal-poor stars with [Fe/H] $< -0.5$.
The shallower slope of the ${<V_{los,GC}}> - l$ relation for their stars
is consistent with the metallicity trends seen in the Minniti and Harding-Morrison
samples. 
 
\section{Comparison with COBE images}

The COBE/DIRBE images in the 1.25, 2.2 and $3.5\mu$m wavelength regions
allow us to compare the distribution of the PNe with the integrated
near-infrared emission from the Galactic bulge. In this region of the
spectrum, the light distribution comes from various stellar populations
but is dominated by the more metal-rich K and M-giants which have
kinematics similar to those of the PNe. The $1.25\mu$m map also gives an
indication of the distribution of the dust.

\subsection{Histogram of the Longitude Distributions}

Figure 14 shows a histogram of the longitude distribution of the COBE light 
and the PNe in our southern surveyed fields ($-5^\circ > b > -10^\circ$). 
The COBE histograms were constructed from the COBE light distribution within 
the individual 30 arcmin fields used for the PNe survey (see Beaulieu et al. 
1999), so the distributions are directly comparable. The dashed lines represents
the three bands  (1.25, 2.2 and $3.5\mu$m) of the COBE light distribution and the 
solid line represents the PNe distribution. We see immediately that the three 
COBE distributions agree very well and that the PNe distribution follows 
the COBE light distribution. The fact that the three COBE light distributions 
agree so well is an indication that extinction, in our surveyed fields, is not 
severe and that its distribution is fairly uniform. 
 
We also compare the three COBE light distributions with their 
cumulative distributions, in preparation for the next section. Figure 15 
shows that the cumulative distributions for the three COBE bands 
are very similar.

\subsection{K-S Test}

The Kolmogorov-Smirnov (K-S) test estimates the probability 
that a set of observed values can be excluded as coming from a given 
specified distribution. 

We performed a one-sample, two-tailed K-S test in our surveyed fields
(Galactic longitude $l = +20^\circ$ to $-20^\circ$ and Galactic latitude
$b = -5^\circ$ to $-10^\circ$) using the well-determined COBE light 
distribution in longitude as the specified distribution and our sample
of PNe as the observed distribution. The test uses the largest value $D$
of the deviation $|F_0(X) - S_N(X)|$ where $|F_0(X)$
and $S_N(X)|$ are the cumulative distributions of the specified
distribution (the COBE light distribution) and the set of observed values
(the longitude distribution of our PNe counts). We have seen in
Figure 15 that the cumulative distributions of the three COBE colors agree
very well and the results for the maximum deviation will 
be similar in all three colors.

Figure 16 shows the two cumulative distributions for our PNe sample and the
$2.2\mu$m COBE light. The ordinate, N, has been normalized to 1.0 for both
distributions. We have used table E of Siegel (1956) to estimate the 
probabilities. Table 6 gives the results obtained for the maximum deviation 
$D$ and the associated probability that the deviation $D$ could occur by chance
from the same parent distribution. Column 1: the COBE band, Column 2: the 
maximum deviation value $D$ and
and Column 3: the associated probability of occurrence. This probability
is between 0.23 and 0.30, and we conclude that there is no significant
difference between the longitude distribution of the PNe and the COBE
light in the zone of our deep survey.

\section{Comparison with models}

The evidence for a bar in our Galaxy initiated much interest in developing
detailed dynamical models (N-body and Schwarzschild) of the Galactic bar-bulge.
Several different kinds of models are now available, but the observational
constraints on their stellar dynamics are still weak. Our kinematical data for
the PNe of the Galactic bar-bulge provide further constraints on the models. 

In this section, we present the data of our survey with velocities relative
to the LSR, using the parameters for the sun's peculiar motion as given in
the equation in \S 4. The motivation for doing so is that most 
observational studies are presented in that manner and it would therefore be 
easier for future comparison. Also, we will use our data to estimate the best
value of the tangential velocity of the LSR for each model. 

\subsection{Presentation of the models}

At the time of conducting this study, there were four triaxial numerical models
available to study the dynamics of the Galactic bulge. They offer interesting
and different approaches to studying the formation and structure of the 
bar-bulge. There are three N-body models (Sellwood 1993; Fux 1996;
Kalnajs 1996) and one Schwarzschild model with an N-body realization (Zhao 1996).  
(Very recently, a more elaborate Schwarzschild model has appeared (H\"{a}fner
et al. 1999), constrained by a subset of the data in Figure 13 plus some
proper motions.)

\subsubsection{Sellwood's model}

Sellwood's model is one of the earlier N-body dynamical models. It is a
purely stellar N-body system with $5 \times 10^4$ particles. It starts
from a Q = 1.2 axisymmetric Kuz'min-Toomre disk 
which contains 70\% of the total mass. The remaining 30\%
is in a rigid Plummer sphere which has half the scale length of the disk.
The bar-bulge forms through the instability of the disk. The resulting 
model shows a peanut-shaped bulge. At a viewing angle of 30$^\circ$
to the major axis and a finite distance from the center, the model shows 
an asymmetry in longitude between
the positive and negative sides, which is consistent with the one seen in 
the COBE/DIRBE image (Weiland et al. 1994).

Figure 17 presents the face-on view (XY) and the edge-on view (YZ) as seen 
from infinity, with the Sun-center line at an angle of 30$^\circ$ 
from the major axis. 

\subsubsection{Fux's model}

Fux's model is an N-body system of stars. It has four components:
an exponential stellar disk of constant thickness ($15 \times 10^5$ particles),
a composite power-law stellar nucleus-spheroid ($5 \times 10^4$ particles),
a dark halo ($2 \times 10^5$ particles), and a dissipative gas 
component (a smoothly truncated Mestel disk with $2 \times 10^4$
particles). The system starts in equilibrium and the rotating bar forms through 
instabilities. The model provided to us by Fux is a gas-free version which has 
evolved for 5 Gyr: we note that Fux (1997) has built more elaborate models
of the Milky Way including gas, which we have not considered here.

Figure 18 presents the face-on view (XY) and the edge-on view (YZ) as
seen from infinity, with the Sun-center line at an angle of 30$^\circ$ 
from the major axis.  

\subsubsection{Kalnajs' model}

Kalnajs has been conducting numerical experiments on thin
self-gravitating disks which turn into triaxial rotating objects because of
buckling instabilities. The projected shapes of these objects, when viewed
from the right distance and orientation, resemble the light
distribution of the Galactic bulge, and the line-of-sight velocities can be
scaled to match observed motions of planetary nebulae in the bulge. The
experiments use only 8000 particles, but since the triaxial objects appear
to be stationary in a rotating frame, one can add the distributions at
different times and obtain models containing effectively $\approx 10^5$
particles.

Figure 19 presents the face-on view (XY) and edge-on view (YZ) as
seen from infinity, with the Sun-center line at an angle of 45$^\circ$ 
from the major axis. 

\subsubsection{Zhao's model}

The last is a model of the COBE bar,
constructed from 10K orbits (direct, retrograde and chaotic)
in the rotating bar potential plus a rigid Miyamoto-Nagai disk potential,
using the non-negative least square fitting technique pioneered by Schwarzschild.

The model provided by Zhao for our comparison is the system allowed to evolve
as an N-body system after 10 rotations and it contains 32634 particles.

Figure 20 presents the face-on view (XY) and the edge-on view (YZ) as
seen from infinity, with the Sun-center line at an angle of 20$^\circ$ 
from the major axis.. 

Table 7 summarizes the parameters suggested by the authors of each model.
Column 1: the model, Column 2: the total number of particles in the
model, Column 3: the solar galactocentric radius $R_\circ$ (in model units), 
Column 4: the viewing angle $\phi$ (in degrees) of the bar. $\phi$ is the angle
between the major axis of the bar and the Sun-center line, and is taken as 
positive for a bar pointing into the first quadrant of l, Column 5: the velocity
scale $V_{scale}$ (km s$^{-1}$ per model unit) of the model, and Column 6:
the solar tangential velocity $V_{\circ,T}$ (km s$^{-1}$).

In the next section, we use a statistical technique to estimate these scaling 
parameters for each model from our data.

\subsection{Search for best parameters}

The authors of each model have suggested values for the Sun's galactocentric
distance (in model units), the viewing angle of the bar and a velocity scale
(Table 7).  
However, by varying these parameters, we may hope to obtain somewhat better
fits to the present data. There are four parameters one can vary:
(i)~the overall spatial scale of the model, or equivalently $R_0$ in
model units; (ii)~the overall velocity scale; (iii)~the viewing angle
of the bar; and (iv)~the tangential velocity of the LSR. Saha (1998)
has developed a method for searching the space of these four
parameters for values which are most likely to have given rise to the
observed data. We used his code, which gives a median fit for the
four parameters and error bar estimates under the assumption that the
models and the data are drawn from the same underlying distribution
function.

We are going to compare the positions and the radial velocities of the 97 PNe 
from the {\it Survey fields only} sample with those of the four models.
We choose to restrict the comparison to the 97 PNe in our survey region,
because they were selected in a homogeneous manner.
In making our comparison we must only use that part of the model which
would fall into our surveyed window. Since our window lies several scale lengths
below the Galactic plane, only a small fraction of the model particles are
used in the comparison. The number of model particles is held fixed as the
observer's position changes: the respective numbers for Sellwood, Fux, Kalnajs
and Zhao were 400, 6000, 9000 and 1700.
  
We use Saha's procedure to make a quantitative comparison
of the $(l,b,V_{los})$ distributions for samples of observed objects and
N-body models. Saha's statistic is

$$W = \prod_{i=1}^{B} \frac{(m_{i}~ + s_{i})!}{m_{i}!~ s_{i}!}$$

\noindent
where the $(l,b,V_{los})$ space has been partitioned into a total
of $B$ cells, $m_i$ and $s_i$ are the numbers of model and sample objects
in the $i$-th cell; $W$ is proportional to the probability
that both the observed sample and the model come from the same underlying
(but unknown) distribution, so $W$ can be used to compare the goodness of fit of 
various models. As described above, the $W$ statistic also serves to estimate the
scaling parameters for each model from the observed sample. (see Sevenster
et al. 1999 for a previous application of this statistic.) 

For choosing the number $B$ of cells, our guideline is that the average
number of model particles per cell should be 5 or more, and the spatial
cells should not be smaller than important features in the distribution
function, such as the scale height (see Saha 1998 for more discussion).
After some experimentations, we used a total of 260 cells in $(l,b,V_{los})$:
13 in $l$, 2 in $b$ and 10 in $V_{los}$. Table 8 presents
the results: Column 1: the model, Column 2: the total number of particles in
our window, Columns 3: the four parameters (i) the solar orbit radius $R_\circ$
in model units, (ii) the orientation angle $\phi$ in degrees, (iii) the velocity 
scale (in km s$^{-1}$ per model unit), and (iv) the solar tangential velocity
(in km s$^{-1}$), Column 4: give the median and the 90 \% confidence limits
for these parameters. 

These results are produced by the program after searching
through the region of parameter space given by $7 < R_\circ < 9$, 
$0^\circ < \phi < 90^\circ$, $200 < V_{\circ,T} < 240$ and
$0 < V_{scale}/V_{scale,model} < 2$, where $V_{scale,model}$ is
the suggested velocity scale value from each model (see table 7).
This choice of search region was partly guided by the likely
values of the corresponding galactic parameters and appears to be
satisfactory: for every parameter and every model, the median estimate 
of the parameter lies away from the boundary of the search region by
at least the 90\% confidence limit. Of these parameters, $V_{scale}$ is 
the least constrained by the data, and $V_{\circ,T}$
the best constrained.  For all of the models, the $W$ statistic indicates that 
the probability that the models and data come from the same underlying 
distribution exceeds 98\%.

\subsection{Models versus Data}

With the estimated parameters given in Table 8, we now present some visual
comparisons of the kinematics of the models and the data. The figures are
similar to those shown earlier for our data alone, except for the fact that
the PNe velocities and the velocity data for the models are heliocentric.

Figures 21 to 24 present the longitude-velocity diagrams for Sellwood
(400 particles), Fux (6000 particles), Kalnajs (9000 particles) and Zhao 
(1700 particles) respectively.

Figures 25 to 28 show $<V_{los}>$ (top panel) and $\sigma$ (lower panel) against 
the longitude for the data and models, with the model represented by thick lines.

The main features of the ($<V_{los}>$,$\sigma$) versus longitude relations are that
all models give a fair representation of the observed  $<{V_{los}}> - l$
distribution, but the Sellwood and Zhao models have a velocity dispersion
that is relatively low. The $V / \sigma$ values for the Sellwood and Zhao
models appear to be somewhat higher than for the bulge of the Galaxy, at least in 
the region of our survey. But, as indicated by the Saha procedure, all of the models
are good representations of the PNe data.

\subsection{Models versus Models}

We attempted to use the program to discriminate between the models by
intercomparing the maximum $W$ value from Saha's procedure for samples of 
similar total numbers
of particles. For example, Sellwood's model has 400 particles in our
survey region, so we estimated values of $W$ for Sellwood's model and
random samples of 400 particles drawn from the larger simulations
(Zhao, Fux, Kalnajs) within our region. Similarly, Zhao's model has
1700 particles within our region, so we compared W values for Zhao's
model and random samples of 1700 particles from the larger simulations
(Fux, Kalnajs). Table 9 presents values of $\ln W$ for each
set of comparisons.

The total number of particles is shown for each comparison. The
sampling standard deviation of $\ln W$ is derived empirically by
the program. Table 9 shows: Column 1: the models
being compared and Column 2: the value of $\ln W$ for each set of comparisons,
i.e. 400, 1700 and 6000 particles. The last line of Table 9 shows the sampling
standard deviation of $\ln W$ of each run. We recall that for all of the models,
the probability that the models and the data come from the same underlying
distribution is more than 98 \%.
 
We see that the values of
$\ln W$ for each model do not differ by more than about
$1.9 \sigma$, indicating again that there is no significant difference
between the ability of the various N-body models to represent our data.
Table 9 shows that Sellwood's model comes out best in the N=400 comparison
of the four models, despite the apparently large deviations in the velocity 
dispersion (Figure 25). We recall that the $W$-statistic involves comparison 
of data and model over cells in velocity {\it and} $(l,b)$. The quality
of the velocity comparisons is seen in Figures 25-28. Figure 29 shows the
cumulative distributions over $l$ of the four models (all with N=400) and
the survey fields only PNe sample (over the same interval in $b$). Sellwood's 
model lies closest to the data in Figure 29, followed by Zhao's model. This
help to understand the ordering of the $\ln W$ values for the models as given
in Table 9.

\section{Summary and Conclusions}

Planetary Nebul\ae~ are good tracers for a dynamical study of the Galactic 
bulge because they are less affected by metallicity
bias than most other tracers and they are strong emitters in H$\alpha$ -
this make their velocities easy to measure. We chose to survey the southern 
Galactic bulge in the region $l = \pm 20^\circ$ and $b = -5^\circ$ to $-10^\circ$
because of its lower extinction relative to the northern bulge.

We compared the longitude distribution of PNe in our surveyed fields
with the COBE light distribution at 1.25, 2.2 and $3.5\mu$m. 
We conclude that (i) the light distributions in the three COBE bands 
agree very well, indicating that the extinction in 
our surveyed fields is not severe and that its distribution is fairly uniform
and, (ii) there is no significant difference between the longitude
distribution of the PNe and the COBE light in the zone of our deep survey.

Recent studies of stellar kinematics in a few clear windows in the Galactic bulge 
have provided mean velocities and velocity dispersions which can be compared with
our data. We thus compared data from Minniti (1996a), Harding and Morrison (1993),
Terndrup et al. (1995) and Ibata and Gilmore (1995a,b), and found that the 
metal-rich stars in Minniti's 
three fields agree very well with our data. Harding and Morrison's metal-rich 
stars agree well with our velocity dispersion data for the PNe, but not so well with
our mean velocity. We also found that the velocity dispersion in Baade's Window
(Terndrup et al. 1995) is somewhat higher than ours near $l = 0$ but note that 
Baade's window is closer to the Galactic plane than most of our PNe.
For the Ibata and Gilmore data, the velocity gradient over $l$ is shallower
than for the PNe and other samples of giants; this is presumably due to their
restriction to more metal-weak giants ([Fe/H]$<-0.5$). Their velocity
dispersion estimates for an isotropic bulge agree better with the PNe values.
 
To assist in the comparison of the four N-body models with our sample of data,
we used a procedure proposed by Saha (1998) to make a quantitative comparison
of the $(l,b,V_{los})$ distributions for samples of observed objects and
N-body models. The main conclusion from this comparison is that all four models
show a fairly good fit to our data.

Sellwood, Fux and Kalnajs' models are all bar-forming systems via the
instabilities of a disk and, after scaling, are kinematically
more or less similar. Zhao's model is constructed to fit the COBE light:
in this sense, it is a step up from Kent's (1992) axisymmetric model for the
Spacelab near-IR photometry. Kent's predicted velocity dispersion, as
quoted by de Zeeuw (1993), was already a fairly good fit to the existing data;
therefore, it is not surprising that Zhao's model should also fit well.

Using the estimated parameters obtained from Saha's procedure, we made some
visual comparisons of the kinematics of the models and the data. The Kalnajs
and Fux models give a good visual representation of the mean velocity and
velocity dispersion of the bulge in our survey region; the Sellwood and
Zhao models represent the mean velocity well but their velocity dispersion 
is marginally low relative to the PNe observations.

It will be interesting to use our PNe sample as a more detailed 
kinematical test of Kuijken's axisymmetric isotropic two-integral model.
One important goal of this comparison would be to look for kinematical
disagreements between the data and the axisymmetric model that might be
kinematical signatures of triaxiality. In the same spirit, it would be
interesting to compare  Kuijken's model in detail with the numerical 
triaxial systems discussed in \S 6.

We saw earlier that Minniti's data (Minniti et al. 1992) is apparently 
not consistent with Kuijken's model. As a preliminary comparison with Kuijken's 
model, we examined the distribution of LSR velocities for our {\it Survey fields}
+ {\it Catalogue} sample. Figure 30 shows a histogram of LSR velocities for
the PNe with $5^{\circ} < l < 10^{\circ}$. We can compare the velocity distribution 
in the region $5^{\circ} < l < 10^{\circ}$ with the distribution measured by
Minniti et al (1992) for the giants towards $l = 8^\circ, b=7^\circ$ and discussed
by Kuijken (1995). The PNe in our $5^{\circ} < l < 10^{\circ}$ region cover a
larger region of sky than the Minniti sample; however, the mean value of $|b|$ for 
the PNe is about $6^\circ$, so we might expect the velocity distributions 
of the PNe and the giants to be at least qualitatively similar. We see from
Figure 30 that the velocity distribution in the region $5^{\circ} < l < 10^{\circ}$ 
is asymmetric; the asymmetry is in the opposite sense to that 
found by Minniti et al. (1992) but closer to that seen for the more
metal-rich stars in Minniti's later (1996b) study for this field.
The mean  LSR velocity for our sample is $36 \pm 11$ km s$^{-1}$, 
compared with $5 \pm 10$ km s$^{-1}$ for the Minniti et al. (1992) sample and
the predicted value of 32 km s$^{-1}$ for the Kuijken model. 
There seems to be better agreement between the PNe and the
two-integral model in this region than was found between the
model and the giants.

It may be that we are seeing an effect of metallicity in the Minniti et al.
(1992) sample. There was no information about metallicity at that time 
so the sample could be suffering from pollution by the more slowly rotating 
metal-poor stars. We recall here that the metal-rich stars in Minniti's three
fields (Minniti 1996a) are in good agreement with our {\it Survey fields} + 
{\it Catalogue} sample (cf Figure 13).

So far, only a few clear Galactic bulge windows have been extensively studied. 
Although these studies provide important information on the kinematics in the
bulge, their small region do not give us the entire picture of the bulge 
kinematics. Two major studies of tracers in the Galactic bulge, the K giants
(Harding \& Morrison 1993) and the OH/IR stars (Sevenster et al. 1997a,b)
and a new PNe H$\alpha$ survey of the Southern Galactic Plane (Parker \& Phillips 
1998) are presently under way. A comparison of the PNe surveys with the results
coming from the OH/IR and K giants large-scale surveys should clearly indicate
any dynamical differences between the populations from which these different 
tracers come.
 
Finally, we conclude that the existing studies give a more or less consistent
picture of the kinematics of the Galactic bulge, as summarized in Figure 13,
at least for the metal-rich bulge tracers. We find it interesting that the
N-body models, in which the bar/bulge grows from the disk via
bar-forming instabilities, give a good representation of the detailed stellar
kinematics of the bulge.
   
\acknowledgments
SFB wish to acknowledge funding from The Australian Government 
through an Australian National University Scholarship and an
Overseas Postgraduate Research Scholarship. We are most grateful to 
Roger Fux and Jerry Sellwood for allowing us to use their models. Our thanks 
go to Gerry Gilmore for critical comments on the manuscript.

\clearpage

%FIGURE CAPTIONS

\figcaption[Beaulieu.f1.eps]{Contours of the COBE/DIRBE $2.2\mu$m image. \label{f1}}

\figcaption[Beaulieu.f2.eps]{$(l,b)$ distributions for the Survey fields only sample
(top panel) and the Survey fields + Catalogue sample (lower panel). \label{f2}}
 
\figcaption[Beaulieu.f3.eps]{Longitude-galactocentric velocity diagram for PNe in the 
Survey fields only (top panel) and the Survey fields + Catalogue sample (lower 
panel). \label{f3}} 

\figcaption[Beaulieu.f4.eps]{Mean galactocentric velocities $V$ (top panel) and velocity 
dispersions $\sigma$ (lower panel), versus longitude for PNe in the Survey
fields only. We used 8 bins of equal number of PNe. Each bin contains 12 to 13 
PNe. \label{f4}} 

\figcaption[Beaulieu.f5.eps]{Mean galactocentric velocities $V$ (top panel) and velocity 
dispersions $\sigma$ (lower panel), versus latitude, using 2 bins of equal number
of PNe in the Survey fields only. Each bin contains 48 to 49 PNe. \label{f5}} 

\figcaption[Beaulieu.f6.eps]{Longitude-galactocentric velocity diagrams for PNe in the 
Survey fields only. The top panel is for the bin in latitude from
$-04.9^\circ$ to $-06.5^\circ$ and the lower panel is for the bin in latitude
from $-06.6^\circ$ to $-10.2^\circ$. Each bin has 48 to 49 PNe. \label{f6}} 

\figcaption[Beaulieu.f7.eps]{Mean galactocentric velocities $V$ (top panel) and velocity 
dispersions $\sigma$ (lower panel), versus longitude for PNe in the Survey
fields + Catalogue sample. We used 12 bins of equal number of PNe. Each bin
contains 31 to 32 PNe. \label{f7}} 

\figcaption[Beaulieu.f8.eps]{Mean galactocentric velocities $V$ (top panel) and velocity 
dispersions $\sigma$ (lower panel), versus latitude, using 6 bins of equal
number of PNe in the Survey fields + Catalogue sample. Each bin contains 62
to 63 PNe. \label{f8}} 

\figcaption[Beaulieu.f9.eps]{Longitude-galactocentric velocity diagrams for PNe in the 
Survey fields + Catalogue sample. The top panel is for the bin in latitude
from $-04.9^\circ$ to $-06.5^\circ$ and the lower panel is for the bin in
latitude from $-06.6^\circ$ to $-10.2^\circ$. The lower latitude bin
(b = $-04.9^\circ$ to $-06.5^\circ$) contains 66 PNe and the higher latitude bin
contains 73 PNe. \label{f9}}
 
\figcaption[Beaulieu.f10.eps]{Longitude-galactocentric velocity diagrams for PNe in the 
Survey fields + Catalogue sample. The top panel is for the bin in latitude from
$+03.3^\circ$ to $+05.2^\circ$ and the lower panel is for the bin in latitude
from $+05.2^\circ$ to $+15.1^\circ$. The bins contains 62 and 63 PNe respectively. \label{f10}} 

\figcaption[Beaulieu.f11.eps]{Longitude-galactocentric velocity diagrams for PNe in the 
Survey fields + Catalogue sample. The top panel is for the bin in latitude from
$-03.3^\circ$ to $-04.4^\circ$ and the lower panel is for the bin in latitude
from $-04.5^\circ$ to $-05.8^\circ$. Each bin contains 62 PNe. \label{f11}} 

\figcaption[Beaulieu.f12.eps]{Longitude-galactocentric velocity diagrams for PNe in the 
Survey fields + Catalogue sample. The top panel is for the bin in latitude
from $-05.8^\circ$ to $-07.4^\circ$ and the lower panel is for the bin in latitude
from $-07.4^\circ$ to $-14.9^\circ$. The bins contains 62 PNe. \label{f12}} 

\figcaption[Beaulieu.f13.eps]{Mean galactocentric velocities $V$ (top panel) and velocity 
dispersions $\sigma$ (lower panel), versus longitude for PNe in the Survey
fields + Catalogue sample. We used 12 bins of equal number of PNe. Overplotted 
are data points from four studies of K-giants: filled symbols for metal-rich
stars and open symbols for metal-poor stars. {\it Circle, star and triangle}:
Minniti (1996); {\it diamond}: Harding \& Morrison (1993); {\it filled square}:
Terndrup (1995); and {\it asterisks}: Ibata \& Gilmore (1995a,b). The line 
represents the slope of the linear rotation curve found for bulge PNe by Durand et
al. (1998) (part of our data is included in their analysis).
More details can be found in Table 5. \label{f13}}

\figcaption[Beaulieu.f14.eps]{Longitude distribution of the COBE light (1.25, 2.2 and 
$3.5\mu$m) and the PNe in the survey fields only. The dashed lines represent the
three COBE bands and the solid line is the PNe. The COBE distributions have
been normalized to give the same area under the histograms as the PNe 
distribution. \label{f14}}

\figcaption[Beaulieu.f15.eps]{Longitude cumulative distribution of the COBE light 
(1.25, 2.2 and $3.5\mu$m) in the survey fields only. \label{f15}}

\figcaption[Beaulieu.f16.eps]{Longitude cumulative distribution of the COBE light at 
$2.2\mu$m and the PNe in the survey fields only. The solid line is the COBE
light cumulative distribution and the {\it staircase} line represent the PNe
cumulative distribution. The ordinate, N, has been normalized to 1.0 for both
distributions. \label{f16}}

\figcaption[Beaulieu.f17.eps]{Face-on view (XY) and edge-on view (YZ) of Sellwood's 
Model. The Sun has been positioned at (X,Y) = (6,0) and is at 30$^\circ$ angle
from the major axis of the bar. The edge-on view point is at infinity. \label{f17}} 

\figcaption[Beaulieu.f18.eps]{Face-on view (XY) and edge-on view (YZ) of Fux's Model. 
The Sun has been positioned at (X,Y) = (8,0) and is at 30$^\circ$ angle from
the major axis of the bar. The edge-on view point is at infinity. \label{f18}} 

\figcaption[Beaulieu.f19.eps]{Face-on view (XY) and edge-on view (YZ) of Kalnajs' 
Model. The Sun has been positioned at (X,Y) = (8,0) and is at 45$^\circ$ angle from 
the major axis of the bar. The edge-on view point is at infinity. \label{f19}}

\figcaption[Beaulieu.f20.eps]{Face-on view (XY) and edge-on view (YZ) of Zhao's 
Model. The Sun has been positioned at (X,Y) = (8,0) and is at 20$^\circ$ angle from
the major axis of the bar. The edge-on view point is at infinity. \label{f20}}

\figcaption[Beaulieu.f21.eps]{Longitude-velocity diagram for Sellwood's model 
with 400 particles. The velocities in this and all following figures are
relative to the LSR. \label{f21}} 

\figcaption[Beaulieu.f22.eps]{Longitude-velocity diagram for Zhao's model with 1700 
particles. \label{f22}} 

\figcaption[Beaulieu.f23.eps]{Longitude-velocity diagram for Fux's model with 6000 
particles. \label{f23}} 

\figcaption[Beaulieu.f24.eps]{Longitude-velocity diagram for Kalnajs' model with 9000 
particles. \label{f24}} 

\figcaption[Beaulieu.f25.eps]{Mean velocity versus longitude (top panel) and mean dispersion 
versus longitude (lower panel) of the PNe (thin lines), with Kalnajs' model represented 
by thick lines. \label{f25}} 

\figcaption[Beaulieu.f26.eps]{Mean velocity versus longitude (top panel) and mean dispersion
versus longitude (lower panel) of the PNe (thin lines), with Fux's model represented 
by thick lines. \label{f26}} 

\figcaption[Beaulieu.f27.eps]{Mean velocity versus longitude (top panel) and mean dispersion
versus longitude (lower panel) of the PNe (thin lines), with Sellwood's model represented 
by thick lines. \label{f27}} 

\figcaption[Beaulieu.f28.eps]{Mean velocity versus longitude (top panel) and mean dispersion
versus longitude (lower panel) of the PNe (thin lines), with Zhao's model represented 
by thick lines. \label{f28}} 

\figcaption[Beaulieu.f29.eps]{Cumulative distributions over $l$ of the four models 
and the Survey fields only PNe (heavy line) (all models with N=400 and over
the same interval in $b$). \label{f29}}

\figcaption[Beaulieu.f30.eps]{Histogram of the distribution of LSR radial 
velocity for our Survey fields + Catalogue sample in the longitude
interval $5^{\circ} < l < 10^{\circ}$. \label{f30}} 

\clearpage

%TABLES

%\include{table1}

\begin{deluxetable}{ccccc}
\tablewidth{0pt}
\tablecaption{($<V>,\sigma$) versus $l$ for {\it Survey fields only}}
\tablehead{
\colhead{Mean longitude} &
\colhead{$<V>$} &
\colhead{$\sigma$} &
\colhead{$<V>_{error}$} &
\colhead{$\sigma_{error}$}}
\startdata
$15^\circ.3$ & $84.9$ & $42.4$ & $12.8$ & $8.6$ \nl
$8^\circ.9$ & $86.8$ & $89.1$ & $26.9$ & $18.2$ \nl
$4^\circ.5$ & $64.4$ & $61.3$ & $18.5$ & $12.5$ \nl
$1^\circ.5$ & $13.9$ & $60.8$ & $18.3$ & $12.4$ \nl
$-0^\circ.9$ & $20.2$ & $103.9$ & $31.3$ & $21.2$ \nl
$-3^\circ.2$ & $-52.4$ & $98.5$ & $29.7$ & $20.1$ \nl
$-6^\circ.1$ & $-54.3$ & $76.8$ & $23.1$ & $15.7$ \nl
$-13^\circ.8$ & $-79.6$ & $90.4$ & $26.1$ & $17.7$ \nl
\enddata
\end{deluxetable}

\clearpage

\begin{deluxetable}{ccccc}
\tablewidth{0pt}
\tablecaption{($<V>,\sigma$) versus $b$ for {\it Survey fields only}}
\tablehead{
\colhead{Mean latitude} &
\colhead{$<V>$} &
\colhead{$\sigma$} &
\colhead{$<V>_{error}$} &
\colhead{$\sigma_{error}$}}
\startdata
$-5^\circ.7$ & $7.4$ & $97.7$ & $14.3$ & $9.9$ \nl
$-8^\circ.1$ & $11.7$ & $102.5$ & $14.8$ & $10.4$ \nl
\enddata
\end{deluxetable}

\clearpage

\begin{deluxetable}{ccccc}
\tablewidth{0pt}
\tablecaption{($<V>,\sigma$) versus $l$ for {\it Survey fields} + {\it Catalogue}}
\tablehead{
\colhead{Mean longitude} &
\colhead{$<V>$} &
\colhead{$\sigma$} &
\colhead{$<V>_{error}$} &
\colhead{$\sigma_{error}$}}
\startdata
$24^\circ.1$ & $138.2$ & $56.1$ & $10.2$ & $7.1$ \nl
$14^\circ.2$ & $74.8$ & $54.3$ & $9.9$ & $6.9$ \nl
$9^\circ.2$ & $76.5$ & $84.1$ & $15.4$ & $10.7$ \nl
$5^\circ.9$ & $56.8$ & $81.3$ & $14.8$ & $10.3$ \nl
$3^\circ.5$ & $46.8$ & $96.2$ & $17.6$ & $12.2$ \nl
$1^\circ.2$ & $-20.8$ & $74.9$ & $13.7$ & $9.5$ \nl
$-0^\circ.7$ & $17.3$ & $103.5$ & $18.9$ & $13.1$ \nl
$-2^\circ.5$ & $-16.5$ & $107.1$ & $19.6$ & $13.6$ \nl
$-4^\circ.4$ & $-66.9$ & $80.6$ & $14.7$ & $10.2$ \nl
$-8^\circ.9$ & $-68.9$ & $79.0$ & $14.4$ & $10.0$ \nl
$-15^\circ.8$ & $-93.9$ & $62.4$ & $11.4$ & $7.9$ \nl
$-24^\circ.5$ & $-135.1$ & $65.4$ & $11.8$ & $8.2$ \nl
\enddata
\end{deluxetable}

\clearpage

\begin{deluxetable}{ccccc}
\tablewidth{0pt}
\tablecaption{($<V>,\sigma$) versus $b$ for {\it Survey fields} + {\it Catalogue}}
\tablehead{
\colhead{Mean latitude} &
\colhead{$<V>$} &
\colhead{$\sigma$} &
\colhead{$<V>_{error}$} &
\colhead{$\sigma_{error}$}}
\startdata
$8^\circ.2$ & $-13.9$ & $96.7$ & $12.3$ & $8.6$ \nl
$4^\circ.2$ & $-6.7$ & $118.0$ & $15.1$ & $10.6$ \nl
$-3^\circ.8$ & $-6.3$ & $124.9$ & $16.0$ & $11.2$ \nl
$-5^\circ.0$ & $17.9$ & $100.9$ & $12.9$ & $9.1$ \nl
$-6^\circ.6$ & $-4.6$ & $103.7$ & $13.3$ & $9.3$ \nl
$-9^\circ.7$ & $15.7$ & $122.5$ & $15.7$ & $11.0$ \nl
\enddata
\end{deluxetable}

\clearpage

\begin{deluxetable}{lcc}
\tablewidth{0pt}
\tablecaption{Comparison Studies}
\tablehead{
\colhead{Study} &
\colhead{region ($l,b$)} &
\colhead{symbol}}
\startdata
Minniti (1996a)\tablenotemark{a} & ($8^\circ$,$7^\circ$) & $\triangle$ \nl
& ($12^\circ$,$3^\circ$) & $\circ$ \nl
& ($10^\circ$,$-7.6^\circ$) & $\star$ \nl
Harding and Morrison (1993)\tablenotemark{a} & ($-10^\circ$,$-10^\circ$) & $\Diamond$ \nl
Terndrup et al. (1995) BW & ($1^\circ$,$-3.9^\circ$) & $\bull$ \nl
Ibata \& Gilmore (1995b) & ($-25^\circ$,$-12^\circ$) & $\ast$ \nl
& ($-15^\circ$,$-12^\circ$) & \nl
& ($-5^\circ$,$-12^\circ$) & \nl
\tablenotetext{a}{ filled = metal-rich and open = metal-poor}
\enddata
\end{deluxetable}

\clearpage

\begin{deluxetable}{ccc}
\tablewidth{0pt}
\tablecaption{Kolmogorov-Smirnov test results}
\tablehead{
\colhead{$\lambda$ ($\mu$m)} &
\colhead{$D$} &
\colhead{Probability}}
\startdata
$1.25$ & $0.1053$ & $0.2319$ \nl
$2.2$ & $0.0991$ & $0.2963$ \nl
$3.5$ & $0.1024$ & $0.2612$ \nl
\enddata
\end{deluxetable}

\clearpage

\begin{deluxetable}{lccccc}
\tablewidth{0pt}
\tablecaption{Suggested parameters for each model.}
\tablehead{
\colhead{Model} &
\colhead{Total \# of } &
\colhead{$R_\circ$} &
\colhead{$\phi$} &
\colhead{$V_{scale}$ (km-s$^{-1}$)} &
\colhead{$V_{\circ~ T}$} \\
\colhead{ } &
\colhead{particles} &
\colhead{model units} &
\colhead{degrees} &
\colhead{per model units} &
\colhead{(km-s$^{-1}$)}}
\startdata
Sellwood & $43802$ & $6$ & $30$ & $300.0$ & -- \nl
Zhao & $32634$ & $8$ & $20$ & $291.0$ & $220.0$ \nl
Fux & $200000$ & $8$ & $30$ & $927.5$ & $213.0$ \nl
Kalnajs & $248000$ & $8$ & $45$ & $150.0$ & $215.0$ \nl
\enddata
\end{deluxetable}

\clearpage

\begin{deluxetable}{lccccc}
\tablewidth{0pt}
\tablecaption{Search for the best parameters in models.}
\tablehead{
\colhead{Model} &
\colhead{\# of fitted} &
\colhead{Parameters} &
\colhead{Median} & \\
\colhead{ } &
\colhead{particles} &
\colhead{} &
\colhead{}}
\startdata
Sellwood & $400$ &$R_\circ$ & $7.3_{-0.2}^{+0.6}$ \nl
& & $\phi$ & $13_{-12}^{+26}$ \nl
& & $V_{scale}$ & $297_{-54}^{+78}$ \nl
& & $V_{\circ~ T}$ & $211_{-9}^{+17}$ \nl
Zhao & $1700$ &$R_\circ$ & $7.2_{-0.2}^{+0.8}$ \nl
& & $\phi$ & $27_{-9}^{+44}$ \nl
& & $V_{scale}$ & $218_{-23}^{+41}$ \nl
& & $V_{\circ~ T}$ & $212_{-11}^{+17}$ \nl
Fux & $6000$ &$R_\circ$ & $8.6_{-0.5}^{+0.3}$ \nl
& & $\phi$ & $9_{-6}^{+17}$ \nl
& & $V_{scale}$ & $751_{-93}^{+93}$ \nl
& & $V_{\circ~ T}$ & $212_{-10}^{+15}$ \nl
Kalnajs & $9000$ &$R_\circ$ & $8.1_{-0.6}^{+0.8}$ \nl
& & $\phi$ & $53_{-43}^{+26}$ \nl
& & $V_{scale}$ & $170_{-18}^{+23}$ \nl
& & $V_{\circ~ T}$ & $217_{-13}^{+18}$ \nl
\tablecomments{$R_\circ$ is in model units,
$\phi$ is in degrees, $V_{scale}$ is in $\rm km~s^{-1}$
per model units and $V_{\circ~ T}$ is in $\rm km~s^{-1}$}
\enddata
\end{deluxetable}

\clearpage

\begin{deluxetable}{lccc}
\tablewidth{0pt}
\tablecaption{$\ln W$ comparison for each sets of models.}
\tablehead{
\colhead{Model} &
\colhead{} &
\colhead{$\ln W$} &
\colhead{} \\
\colhead{} &
\colhead{$400$} &
\colhead{$1700$} &
\colhead{$6000$}}
\startdata
Sellwood & $129.6$ & \nodata & \nodata\nl
Zhao & $128.45$ & $239.25$ & \nodata\nl
Fux & $126.67$ & $237.47$ & $352.75$\nl
Kalnajs & $126.16$ & $237.95$ & $350.82$\nl
$\sigma$(sampling) & $1.84$ & $1.89$ & $1.50$ \nl
\enddata
\end{deluxetable}

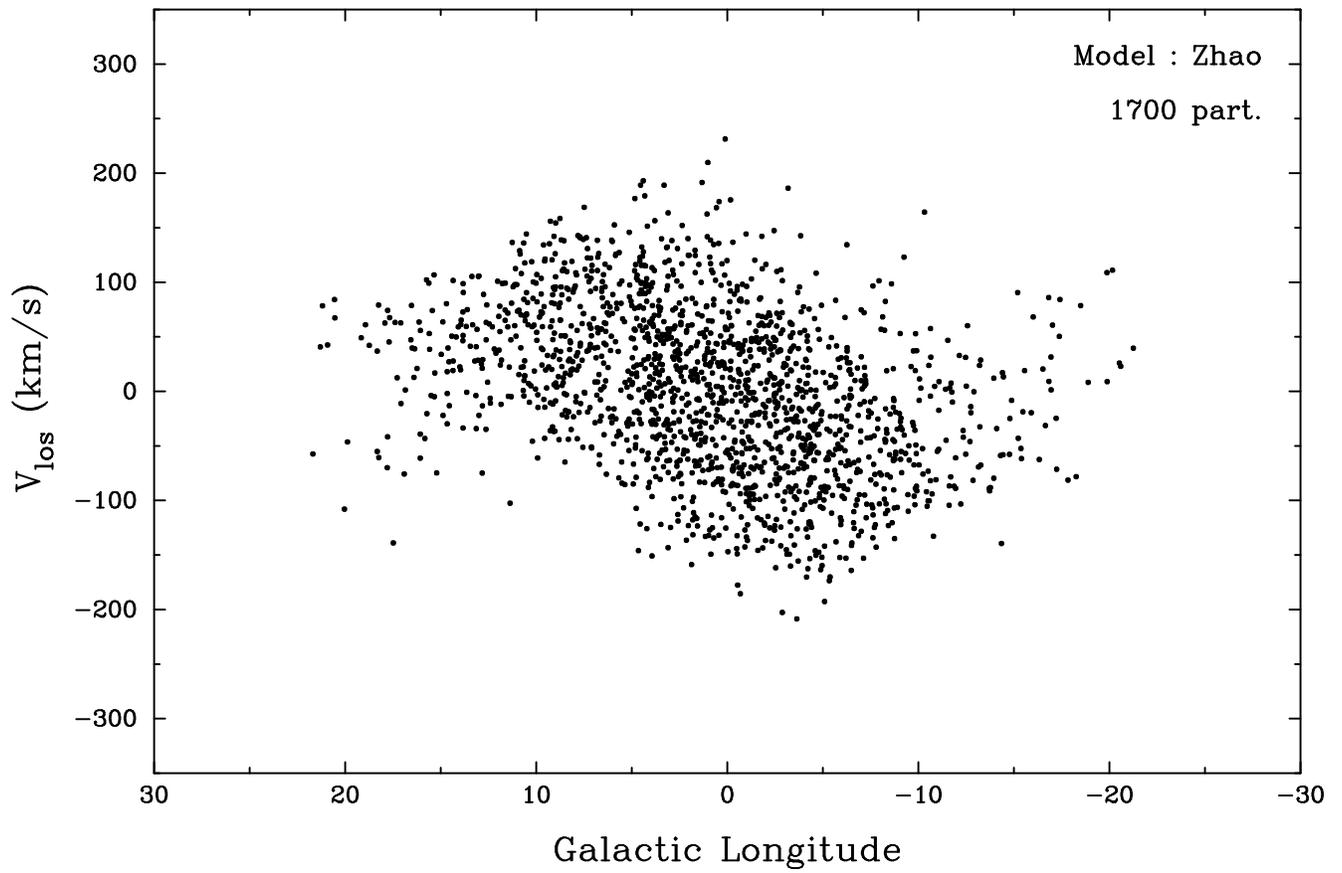
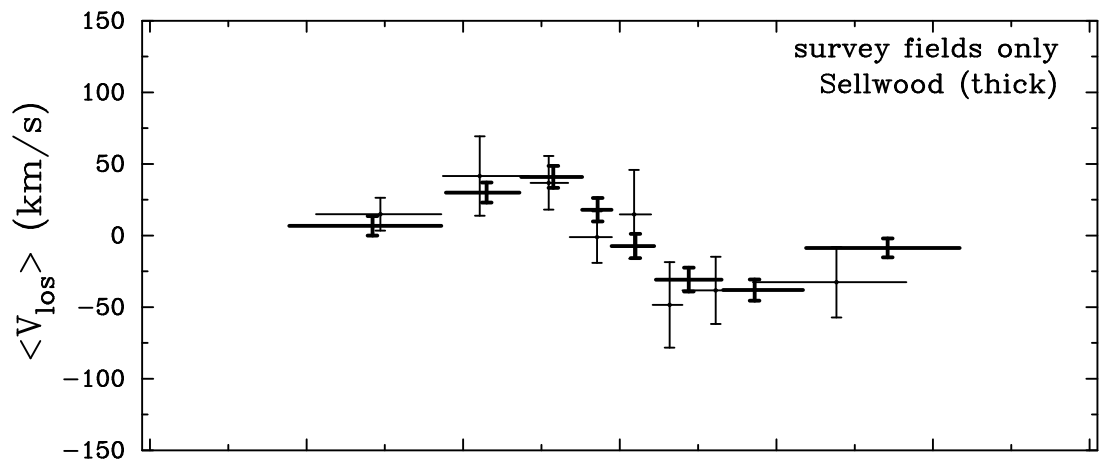
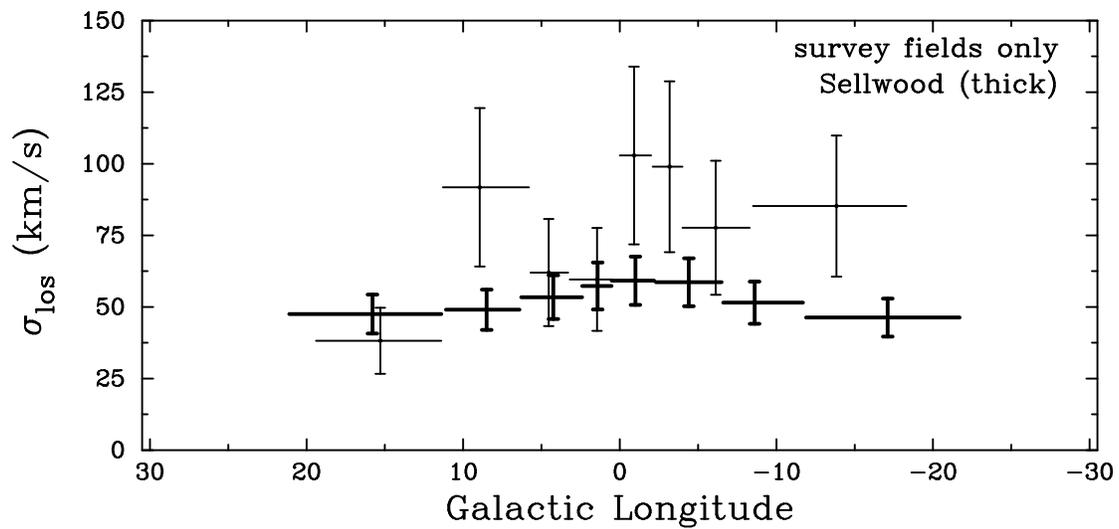
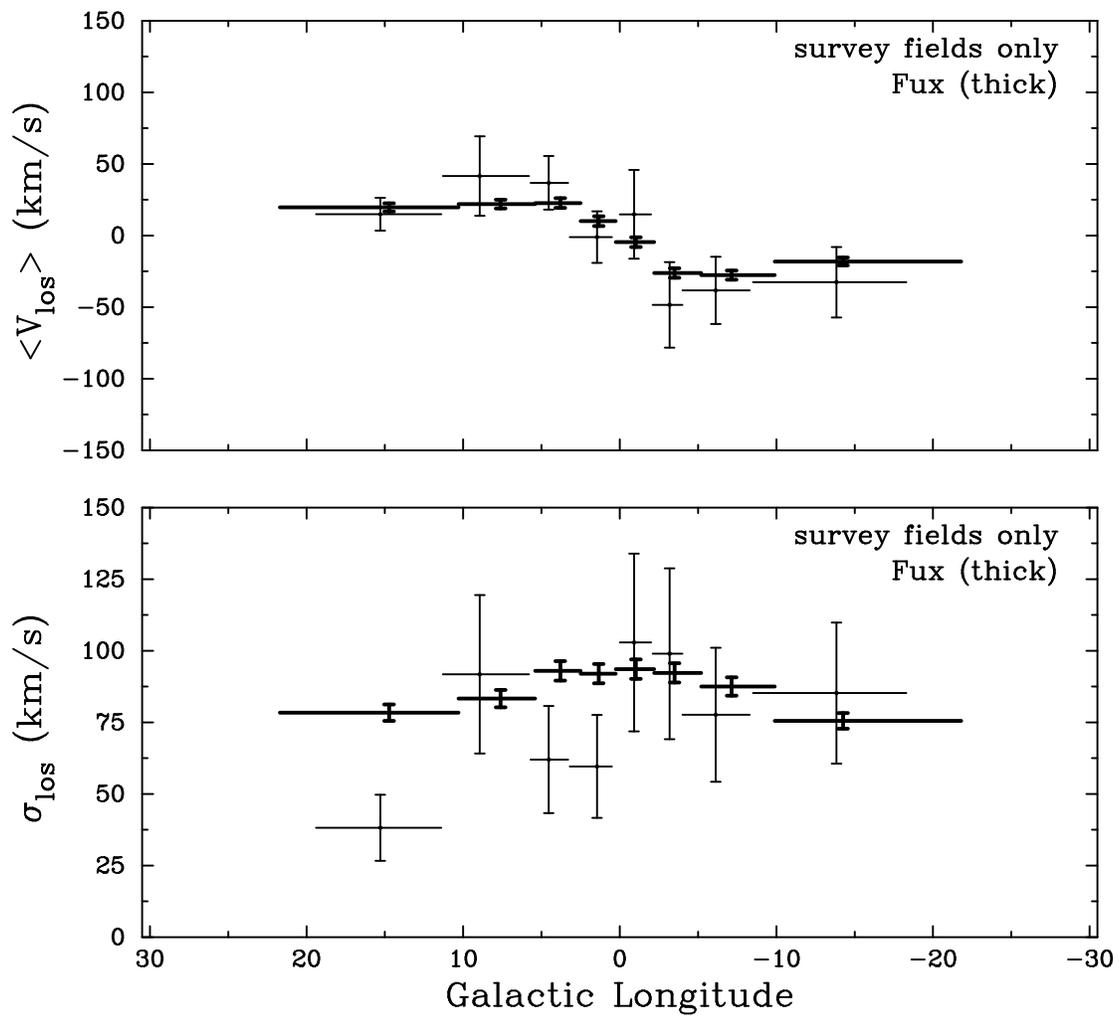
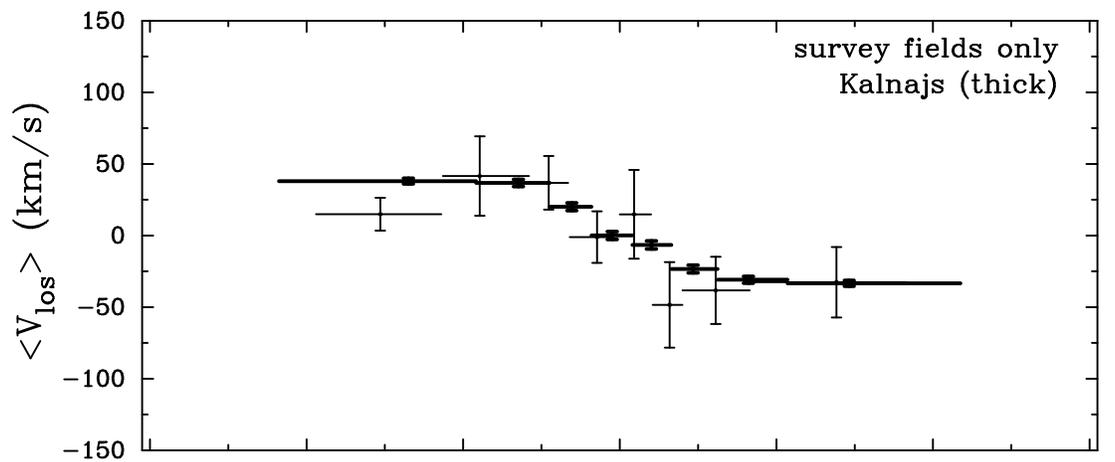
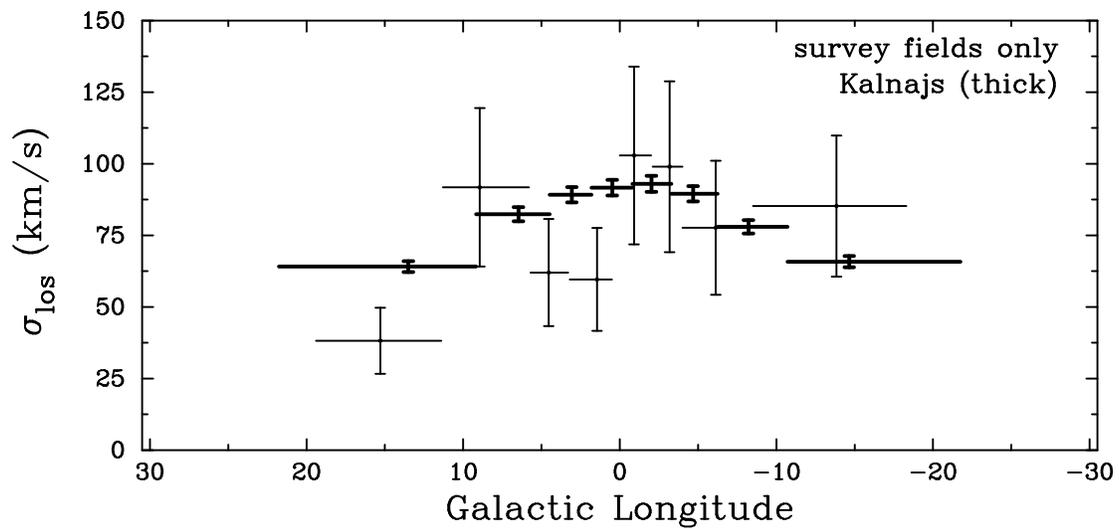
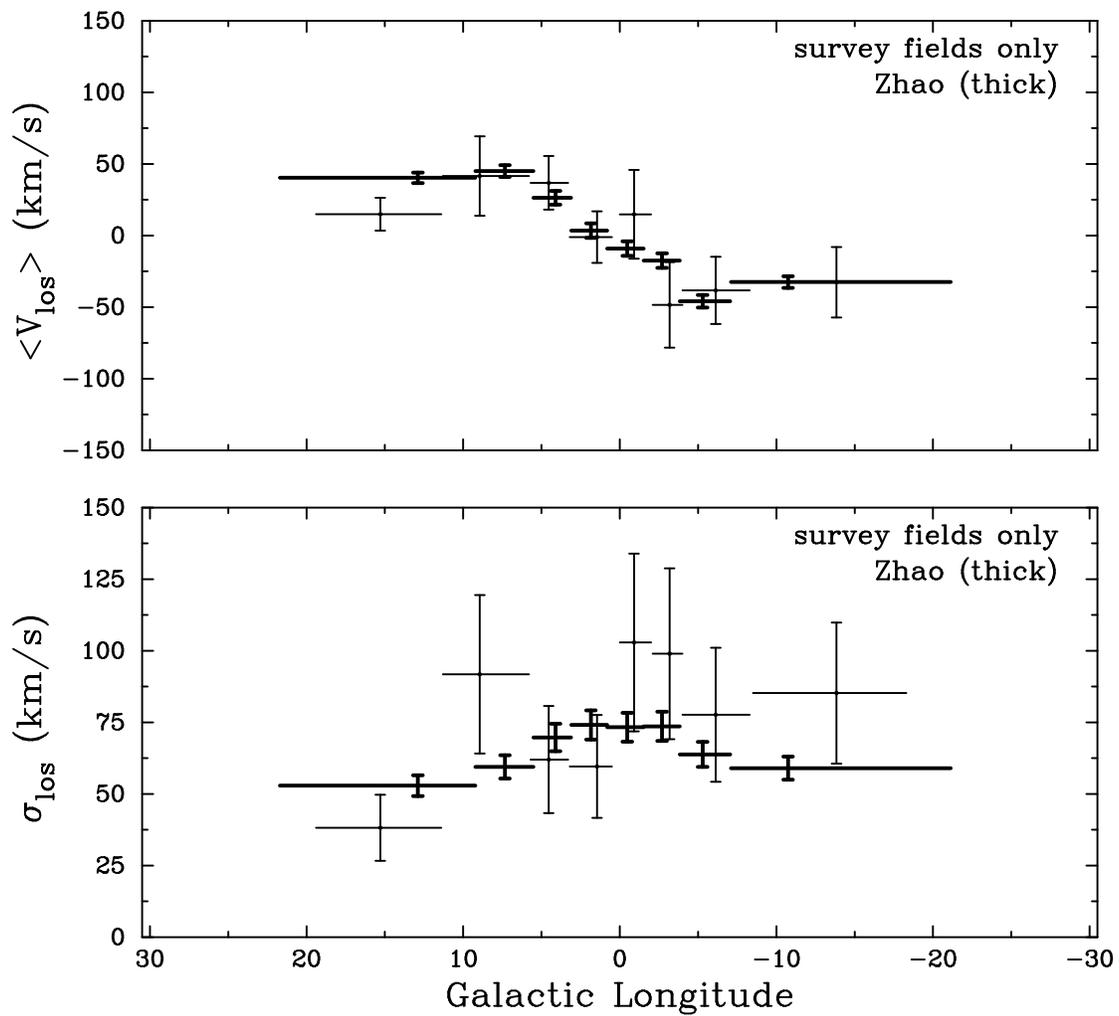
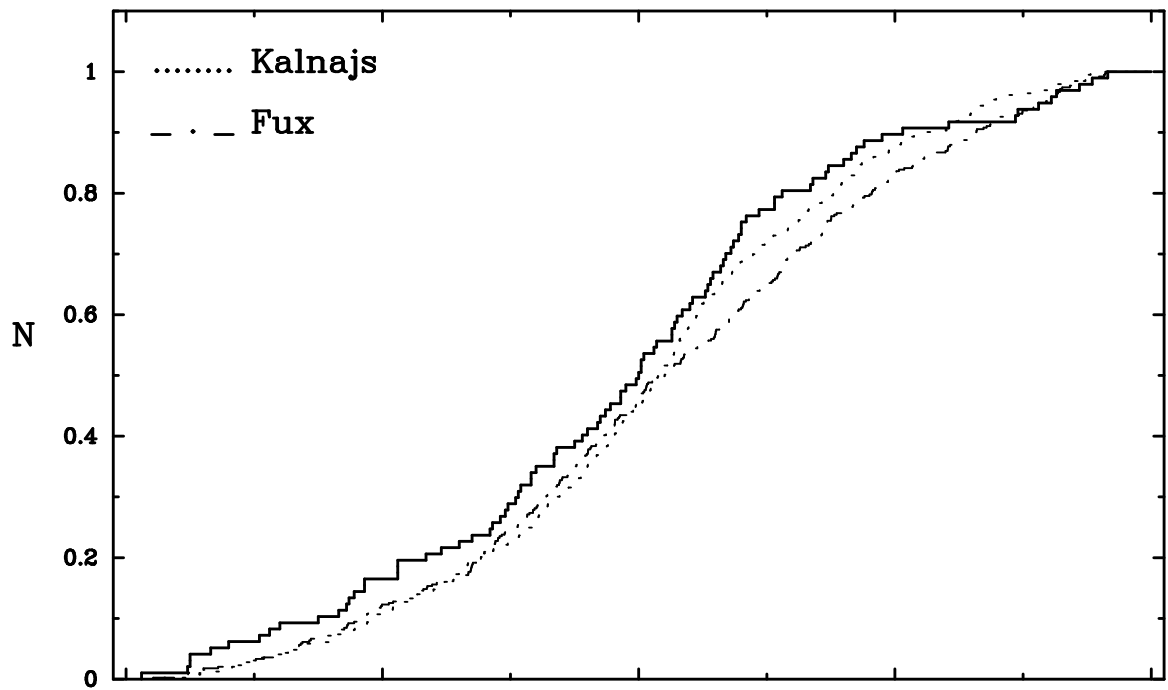
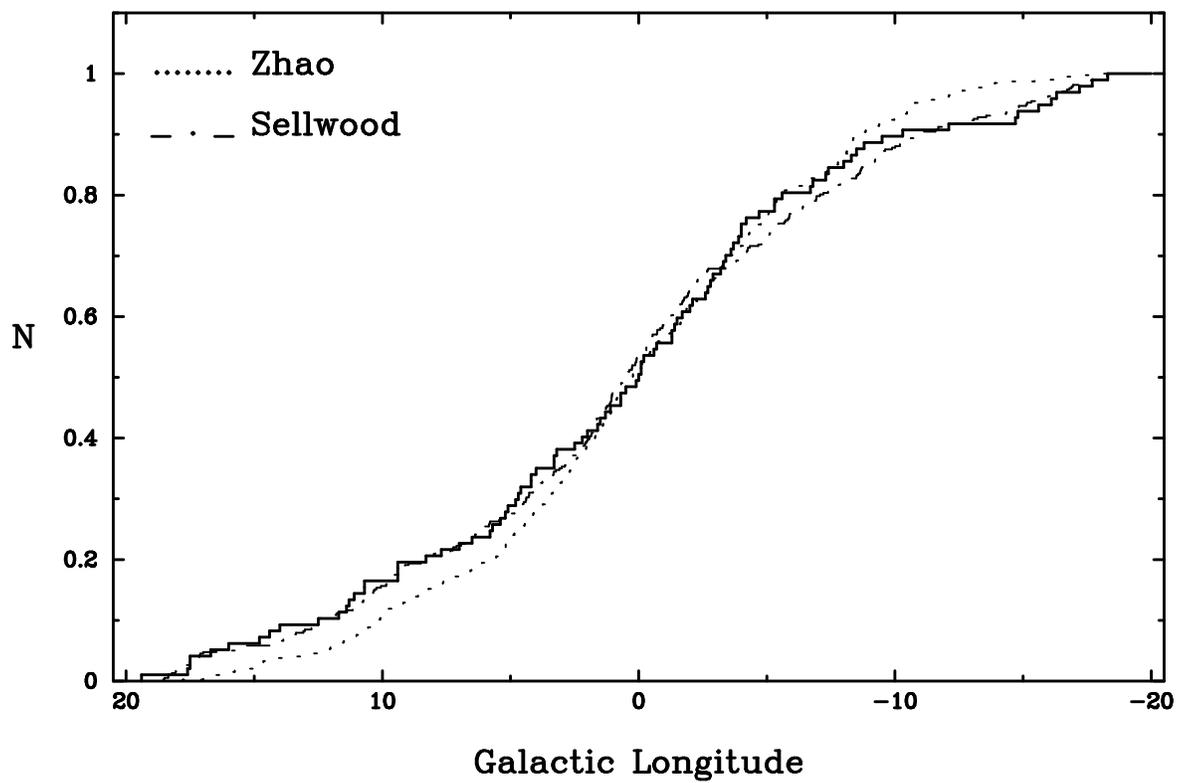
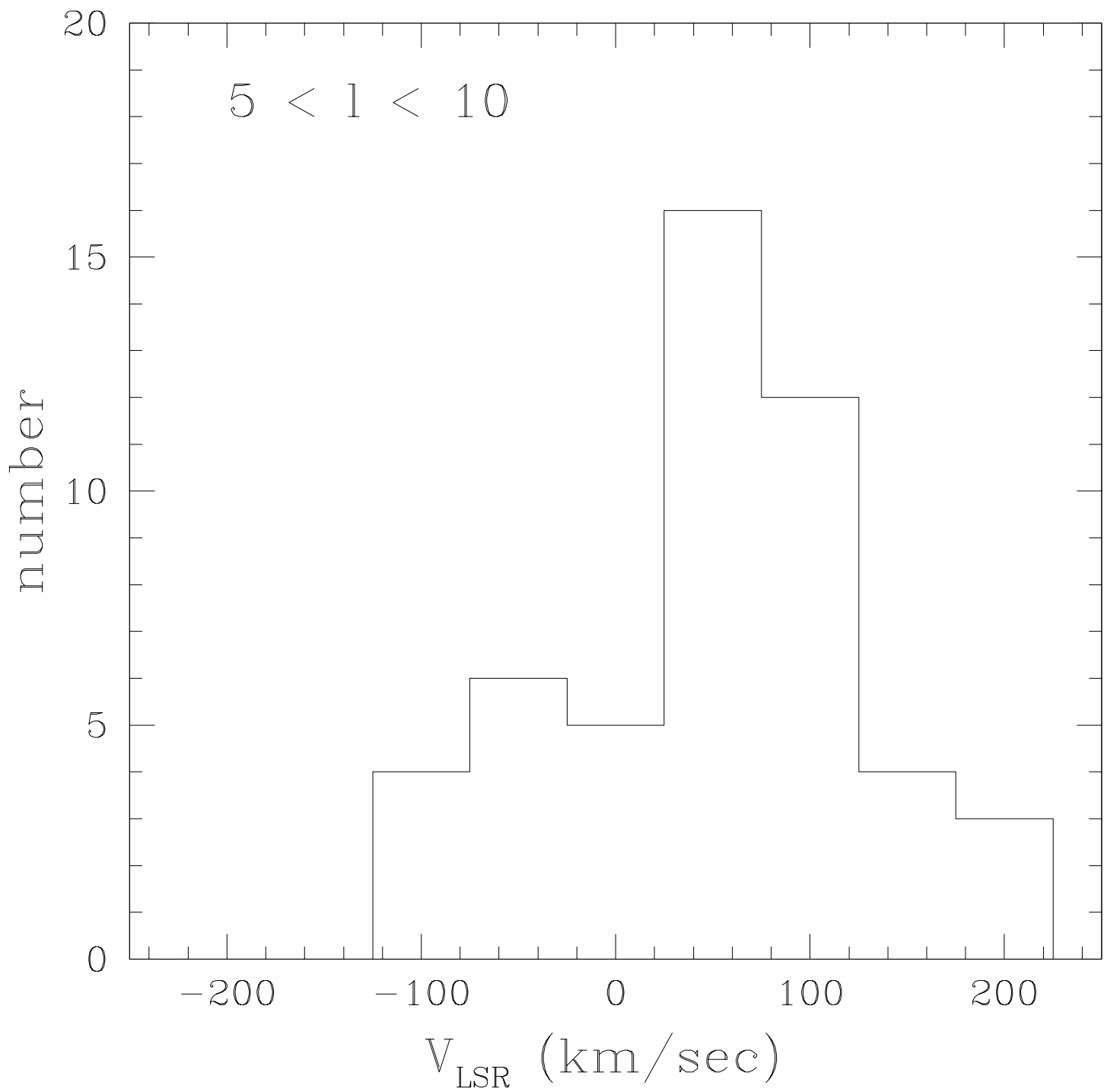
\end{document}